# System level specification and verification using Concurrent State Machines and COSMA environment. A case study.[1]

by

**Wiktor B. Daszczuk, Jerzy Mieścicki, Michał Nowacki, Jacek Wytrębowicz**

wbd@ii.pw.edu.pl , jms@ii.pw.edu.pl , mno@ii.pw.edu.pl , jwt@ii.pw.edu.pl

**Institute of Computer Science**
**Warsaw University of Technology**

ul. Nowowiejska 15/19, 00-665 Warsaw, Poland

**Abstract.** Traffic Light Controller, a typical benchmark device, is specified and verified using of a formal model called Concurrent State Machines (CSM) and the software environment COSMA 2.0, which supports the system level specification and analysis of concurrent, asynchronous and communicating units. The TLC itself is a system of three concurrent components (the controller and two timers). The paper introduces briefly the CSM model and illustrates how system components are specified, how the reachability graph of a system is obtained and how the requirements are formally verified. Finally, the hints for the generation of VHDL code for the TLC are given.

## 1. Introduction

The design of a simple Traffic Light Controller was selected as a teaching example of capabilities of a formal model called Concurrent State Machines (CSM) and of the software environment COSMA 2.0, now under implementation in the Institute of Computer Science, WUT. Although the CSM model itself [6, 7, 8, 13] as well as the COSMA environment [12] are directed mainly towards the design and verification of software [9], the methodology was tested also on the verification of communication protocols [10, 11] and the attempt to apply it to the hardware design seemed interesting.

To the formal verification of hardware a vast body of literature is devoted (see e.g. the reviews in [1, 2, 4]). However, more detailed discussion of formal methods and tools falls far beyond the scope of this paper. It reports just a case study, aimed to demonstrate the convenient, intuitive character of the CSM specification, the applied method of the verification (temporal symbolic model checking), as well as the functionality of several software modules that are parts of COSMA 2.0.

The main advantage of the CSM model is that it allows for the specification and formal verification of the system consisting of several units, which operate and communicate concurrently and asynchronously. Verification may detect the potential harmful errors in the communication and synchronization among system components and their environment, like a deadlock, livelock, possible lack of response for some specific event, unwanted simultaneous activity of two components (e.g. violation of the mutual exclusion requirement) etc. Once such errors are identified and cured, the designer can focus his/her attention on the implementation of individual components or modules of the system.

---

[1] This work was supported by grant No. 503/1915 from the Dean of the Faculty of Electronics and Information Technology, Warsaw University of Technology.

The Traffic Light Controller (TLC) is one of benchmark circuits [3] used to test the functionality and the performance of methods and tools developed for the hardware design purposes. Its natural-language description (in an original form proposed in early 90's by Gupta and Ramachandran) is given in Appendix A. In Section 2 the required functionality of the TLC is discussed. Section 3 briefly introduces the CSM model and contains the CSM specification of the TLC, viewed as a system of three components. Also, the Reachability Graph of a system is given. In Section 4 the functional requirements for the TLC are formally expressed in a form of temporal formulas which consecutively undergo the evaluation in the Reachability Graph. In Section 5 contains the guidelines for the generation of VHDL code from the CSM model of TLC. Final remarks in Section 7 conclude the paper.

**2. Functions of a Traffic Light Controller**

The purpose of the TLC is to control safely and smoothly the traffic in the intersection of the highway and the farm road (Fig. 1). As its output, TLC has to switch on and off two sets of traffic lights: *HR, HY, HG* (for highway Red, Yellow, Green, respectively) and *FR, FY, FG* (for farm road, analogously). The input for TLC is car sensor, indicating the presence of the farmer's car (or cars) approaching the junction on either side of the highway. Normally (i.e. if no cars are present in the farm road) the lights should be green for highway (*HG*) and red (*FR*) for farm road. If the farm car approaches, *HG* light should change to yellow (*HY*) for a 'short' time TS and then become *HR* and *FG* to let the farmer pass. However, the highway traffic can not be stopped for a time longer than ('long') TL, even if the whole train of farm cars waits for the passage. On the other hand, if the farmer passes the intersection quickly, it is reasonable to resume the highway traffic immediately, not 'consuming' the whole 'long' time slice TL. The change from (*HR, FG*) back to (*HG, FR*) goes again through yellow lights in the side road (first *HR, FY*, then *HG, FR*).

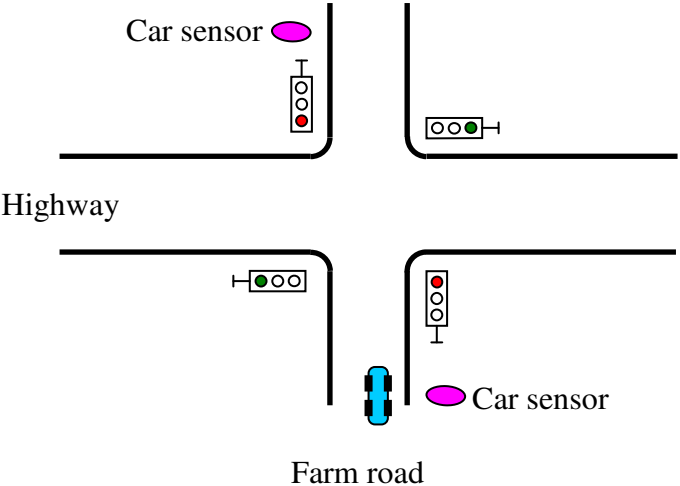

**Fig. 1. The road intersection controlled by the TLC**

The reader is encouraged to read the full original functional requirements, as specified in Appendix A. The essence of the TLC functionality is illustrated in Fig. 2. As can be seen, the controller should have four states and a set of transitions among them, executed when the appropriate condition becomes true. In addition to the controller itself, the TLC system has to provide two timers: one for 'short' time interval (or TS), other for 'long' interval TL. The former (Timer TS) counts just the fixed time interval for the yellow lights be on. The latter (Timer TL) is somewhat more sophisticated one: once started, it determines long timeout TL for green lights in either direction, but it should be *resetable,* in the case the controller wants to quit the FG time slice sooner.

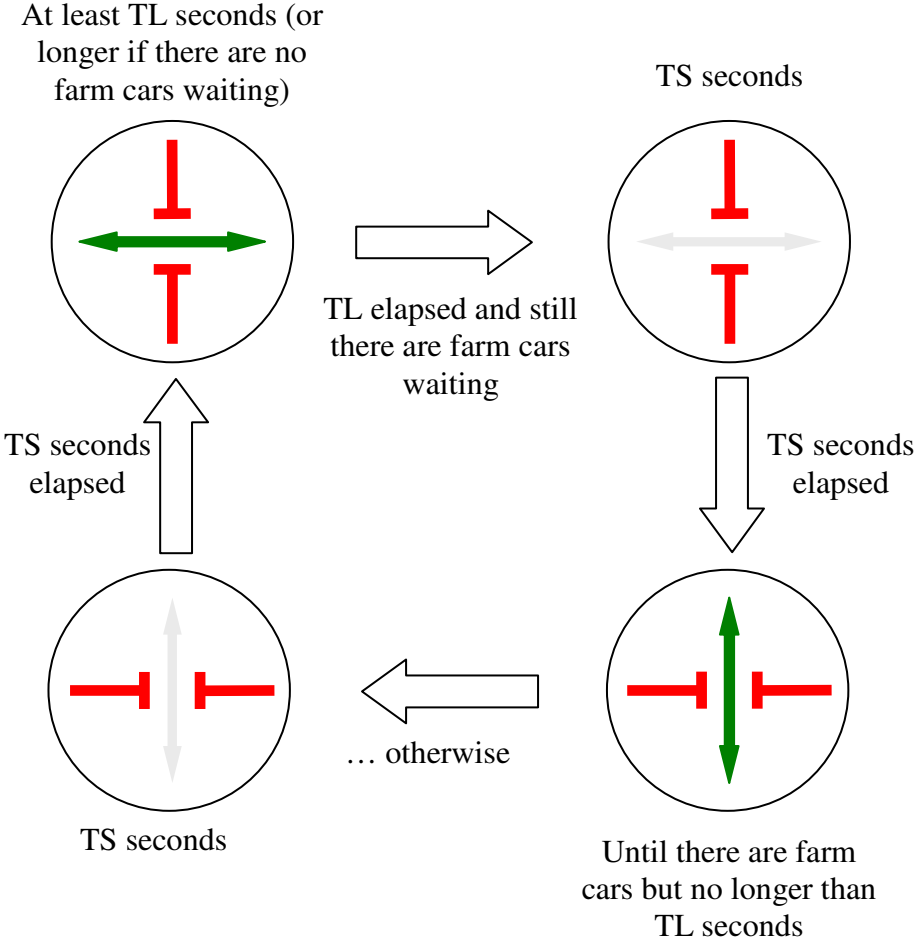

**Fig. 2. The illustration of the TLC functionality**

## 3. CSM specification of the TLC

### 3.1. Brief introduction to Concurrent State Machines

Concurrent State Machines (CSM) are labeled, directed graphs, which represent in an abstract way some discrete objects, e.g. hardware control units, programs, processes, protocols, etc. In other words, they can be used as *formal models* of these software or hardware devices, units, components and even whole systems. The ultimate goal of this modeling is the analysis or verification of the behavior of a system.

Toward this end, one has firstly to develop the structural diagram of the designed device, identifying its basic structural sub-units as well as the communication connections among them. In the case of hardware units it is a typical device's block-diagram where communication links correspond to bus lines or direct signals exchanged among individual blocks. The structural diagram of the TLC (Fig. 3 and Table I) clearly results from the analysis of design requirements and is easily understandable.

Then, the CSM models of individual system components have to be developed. Each of them (see e.g. Figures 4, 5 and 6) at a first glance resembles a typical and well-known Moore finite automaton (or finite state machine, FSM). Indeed, any CSM is a graph, consisting of:

- finite set of *nodes*, interpreted as *states of component's behavior*,
- directed labeled *arcs*, interpreted as elements of the *next-state relation*.

However, in contrast to conventional finite state machines, in CSM arcs are labeled with *Boolean formulas* instead of symbols from an input alphabet. For instance, formula *a* would mean that 'symbol a occurs at machine's input'[2]. Similarly, *b* means that 'the symbol b is present', formula *~a\*~b* means that 'neither a nor b occurs' etc. The arc ($s$, $s'$) from node $s$ to $s'$, labeled with formula *f*, means that $s'$ can follow $s$ if formula *f* is true. In a case when $s = s'$ (i.e. an arc makes a 'loop' over the same state) formula *f* represents a condition under which the machine can *remain* in $s$. Otherwise, i.e. if $s \neq s'$, the arc represents a *transition* (from $s$ to $s'$) while its formula *f* specifies a condition that enables this transition. Note that two or more Boolean formulas can be simultaneously true and – consecutively - more than one arc from a state can simultaneously enabled. Then, *only one* of them is selected. The choice is non-deterministic. Note also that arcs labeled with the condition *1* ('unconditionally true', by the definition) can be used. They are interpreted as *spontaneous transitions* that require no external events or messages to be enabled.

Thus, Concurrent State Machines (as an abstract model) represent the conditions for changes of states in terms of occurrences of abstract symbols from some finite input alphabet. The practical interpretation of these symbols depends on the nature of a system under consideration. In a model of communicating software processes, 'symbols' may stand for specific events, messages or conditions. For instance, the Boolean formula '*e1\*c2 + ack*' would mean that a process has to execute some action 'if event e1 occurs while condition c2 is satisfied or if message ack comes', etc. In hardware models, symbols are usually interpreted in terms of logical values (set - reset, on - off, 0 -1) assumed by binary variables. For example, the formula '*ready\*~bbsy*' would mean that the transition has to be executed 'if bus line ready is set to1 while bbsy is reset to 0'. Note that this use of abstract symbols instead of application-specific conventions is one of important advantages of the CSM model, because it provides the common framework for analysis and design of co-designed hardware/software structures.

The key point in the CSM model is that (again in contrast to conventional FSM) the sequential occurrence of input symbols is neither assumed nor guaranteed. While a FSM receives the neat, purely sequential input tape (or string) of symbols – the CSM machine deals with much less restricted input. Two symbols, a and b, say, are not 'pre-synchronized' (e.g. sequenced or interleaved) in any way. At any instant of time, they can come either alone or simultaneously or even not come at all. Moreover, any component of a system can transmit its own output symbols[3] that can be inputs to neighboring machines (and even to itself). No implicit synchronization among component's activities is assumed[4]. This way, the CSM model supports communication among mutually asynchronous, concurrent system components and their environment.

---

[2] We will use the convention that the symbols themselves are underlined while formulas that refer to them are printed in *italics*.

The key element of the CSM model is the algorithm for computing so-called Reachability Graph (*RG*) of the system. Once CSM models of all components are specified - the algorithm computes all states immediately reachable from system's initial state (along with Boolean formulas that enable the transitions), then - all states reachable from the ones obtained in the first step, then again all states reachable from newly computed states etc., until no new states and transitions occur as the result of the computation. This way we obtain the *product* of individual models of components, showing *all* configurations or co-incidences of their states and *all* transitions that are likely to occur.

The analysis of *RG* may detect and identify unexpected and even harmful synchronization and communication errors, like a deadlock, a livelock, possible lack of response for some specific event, unwanted simultaneous activity of two components (e.g. violation of the mutual exclusion requirement) etc. These errors are practically unavoidable in the design of non-trivial structures involving the asynchronous cooperation among several concurrent units. What even worse, in many cases they are hardly detectable by simulation and testing, as they may result from very rare coincidences of components' states and external stimuli. On the other hand, Reachability Graph includes *all practically possible* states and transitions, therefore it highlights all possible, even very rare, sequences of events which are just paths in the *RG*.

Of course, *RG* can be of an enormous size, which results in well-known time and space complexity problems. In the case of simple systems analyzed just for tutorial purposes (like the one discussed in the present paper), where the *RG* includes only a dozen or two of states, one can draw or print the *RG* and analyze it 'by hand'. In more practical cases, the number of *RG* nodes (i.e. system states) can be of order of $10^{20} - 10^{50}$ or even more [5]. To manage the problem, large graphs are usually represented in a form of data structures known as ROBDD (Reduced Ordered Binary Decision Diagrams [xxx]) that allow for very concise representation. Due to this, in many practical cases the development and analysis of system's RG does not exceed storage and processing power capabilities of an average workstation.

Understandably enough, the inspection of such a large *RG* cannot be done 'by hand' or 'by naked eye'. Thus, one should formally specify the *requirements* for system's behavior and then use the appropriate algorithm for the evaluation if these requirements are actually satisfied in a given *RG*. The commonplace approach involves the use of *temporal logic*, where the requirements have the form of *temporal formulas*. There are many types of temporal logic [xxx], but generally they allow for constructing formal sentences, where temporal connectives (*always*, *eventually*, *next*, *until*) can be used in addition to 'classical' or Boolean operators (*not*, *and*, *or*, *if .. then ..* etc.) and two quantificators (*for all ..*, *exists..*). Temporal propositions expressed this way can cover a very wide class of requirements addressing the issues of the flow of control, communication and synchronization among components as well as other problems of sequencing of events.

---

[3] In a CSM model, outputs are attributed to machine's states, much like to Moore-type FSM. However, unlike to FSM, it is not required that input and output alphabets are disjoint: this way the CSM can transmit output symbols 'to itself'.

[4] Generally, no common clock is assumed, but (if necessary) the clock or even multiple, independent clocks can be introduced in an open way, as simple, two-state CSM automata with spontaneous transitions and the appropriate output symbols which enable relevant transitions in individual components.

The above-described technique for the verification of concurrent systems is referred to as *temporal symbolic model checking* [5]. It is used mainly for the verification of *control-dominated* devices or processes, as it does not support the verification of operations on more general data types. In the latter case, the approach based on *theorem proving* has to be applied rather than the exhaustive inspection of a large (but finite) Reachability Graph of a system. These two general approaches (i.e. finite state models and theorem proving) have their methodologies and supporting tools, along with their numerous specification languages, examples of applications etc. As more detailed discussion of formal methods and tools falls beyond the scope of the present papers, the reader is encouraged to visit http://archive.comlab.ox.ac.uk/formal-methods, where the relevant information is available.

In the Institute of Computer Science (Warsaw University of Technology), an original software tool COSMA 2.0 is now under development. The main part of the present version of COSMA consists of three modules: Grapher, Product Engine and TempoRG. Grapher provides the user interface for drawing and editing CSM models. It also converts graphical specification of system components into XML-like language called CXL. Product Engine converts the CXL specification into a set of Binary Decision Diagrams (BDD) and then computes the system's Reachability Graph, which is again a large BDD. This module uses the state-of-the-art. library of functions for processing ROBDDs, implemented by Geert Janssen from Eindhoven University of Technology [xxx]. TempoRG [xxx] contains a set of algorithms for the evaluation of temporal requirements in a given RG of a system. Additionally, other two modules are available in order to facilitate the verification of behavioral models specified in UML [xxx], which is a specification tool now frequently used by software engineers. One of them supports the conversion of UML state diagrams into CSM models while the other is used for conversion of UML sequence diagrams into temporal requirements.

All these modules make COSMA a powerful symbolic model checker, designed primarily for the verification of synchronization and communication in concurrent software. The unique feature of the approach underlying the COSMA environment is that it is based upon *two* closely related models: CSM (which has been briefly introduced above) and Extended CSM (ECSM). The latter one enhances the expressive power of CSM, as it allows for specification of general data structures and arbitrary operations on these data. Operations on data can be attributed either to states or to transitions of the CSM, which becomes this way just a scheme of flow of control in a process. It is allowed also to replace actual operations by specified distributions of execution time, actual logical conditions - by branching probabilities etc., to enable the performance evaluation by the *simulation* of the system's behavior. The ECSM Simulator is also the part of COSMA 2.0 environment. This way, in the design process of a concurrent, asynchronous system one can verify the correctness of communication and synchronization mechanisms by finite state model checking and evaluate the system's performance by simulation of its ECSM model as well.

### 3.2. TLC structural diagram

The TLC is implemented as a system of three concurrent components (Fig. 3). CONTROLLER makes the 'heart' of the system while two timers (TIMER TS and TIMER TL) provide the indication of time intervals. Both timers are *logical* units, in a sense that their behavior can be expressed in terms of states and transitions labeled with the Boolean formulas. It is assumed that each logical timer has also an own *physical clock* which is somehow able to measure physical time intervals and which sets the Boolean value *tauTS* or *tauTL* (respectively) to *true* when the appropriate time interval ends. The analog circuitry necessary for this purpose is not discussed here. This interpretation is consistent with (and is even more detailed than) the original benchmark description given in Appendix A.

The only input from the environment is the *Car* signal. Signals *HR, HY, HG, FR, FY, FG* make the output of the system. The meaning of remaining signals is explained in Table I.

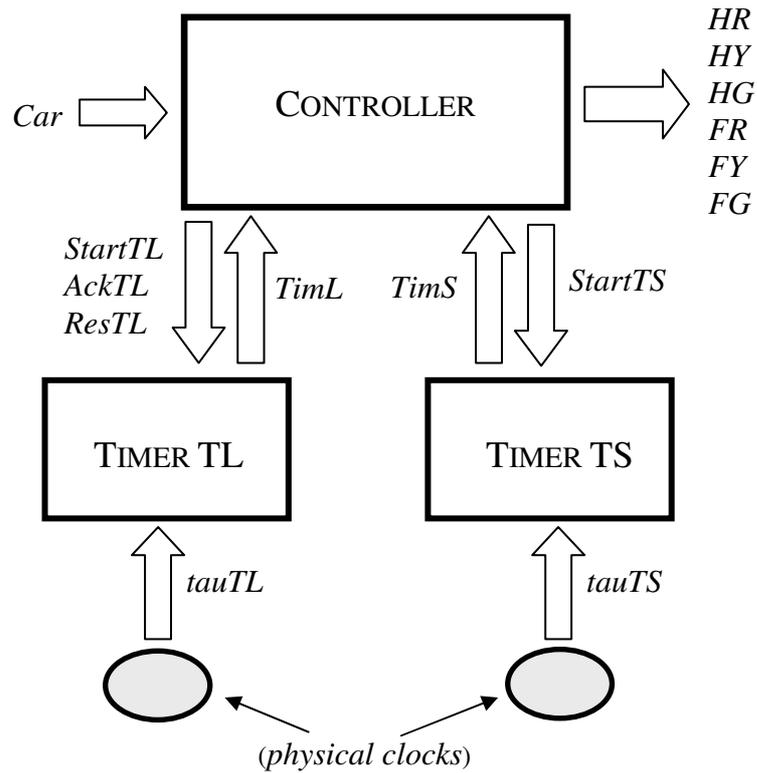

**Fig. 3. Block diagram of the TLC system**

**Table I.**
**Interpretation of signals in the TLC system**

| Signal | Meaning (when *true*) |
|---|---|
| *Car* | Farm car waits to pass |
| *HR, HY, HG* | Red, Yellow, Green light (resp.) for highway is *on* |
| *FR, FY, FG* | Red, Yellow, Green lights (resp.) for farm road is *on* |
| *StartTL* | Start Timer TL |
| *AckTL* | Set Timer TL back to initial state after it has measured the whole TL interval |
| *ResTL* | Reset Timer TL immediately |
| *TimTL* | TL interval elapsed |
| *StartTS* | Start Timer TS |
| *TimTS* | TS interval elapsed |
| *TauTL* | TL time units measured (a logical signal from physical clock) |
| *TauTS* | TS time units measured (a logical signal from physical clock) |

### 3.3. CSM models of system components[5]

### 3.2.1 CSM model of the CONTROLLER

CSM model of the CONTROLLER is shown in Fig. 4. Rounded boxes represent states, directed arcs – transitions. The upper part of the state box contains the state name or identifier while lower part of each state box contains output symbols (signals) produced in this state. Initial state is identified by a thicker borderline of a box (in this case initial state is *s000* ).
 Labels attributed to arcs indicate Boolean conditions that make the transition enabled. 'Self-loop' from the state to itself means that the device *remains* in a given state until the self-loop condition is *true*.

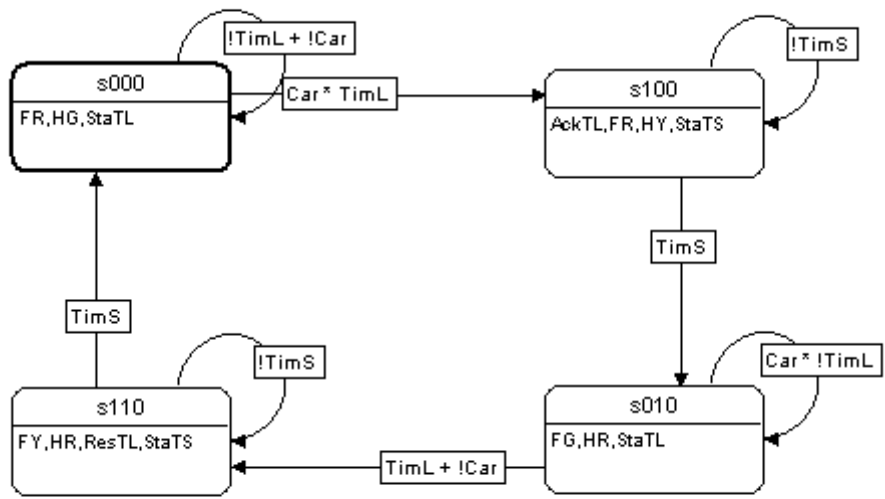

**Fig. 4. The model of the CONTROLLER**

What should be emphasized is a striking similarity of the informal description of TLC behavior from Fig. 2. and the formal CSM model from Fig. 4. Boolean formulas at the graph edges are in fact the straightforward 'translations' of intuitive sentences from Fig. 2. Also the output from the system to traffic lights (*FR, FY, FG, HR, HY, HG*) is easily attributable to individual controller's states.

### 3.2.2 CSM model of the TIMER TS

TIMER TS (Fig. 5.) is in an initial state (*TSidle*) until it receives *startTS* from the CONTROLLER. Then it passes to *TSrun* state where it remains until the Boolean variable *tauTS* (from hypothetical physical clock) becomes *true*. Then it enters TS*elap* state, where the signal

---

[5] All the models discussed in this section were drawn using the Grapher module, which is a part of the COSMA environment responsible for the edition of CSM models.

*TimS* is produced and immediately (unconditionally, with an always *true* condition *1*) it returns back to *TSidle*. Note that neither *TSidle* nor *TSidle* produce any output symbols.

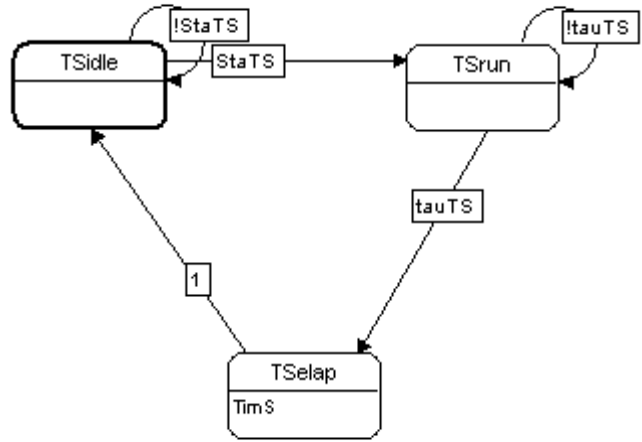

**Fig. 5. The model of the TIMER TS**

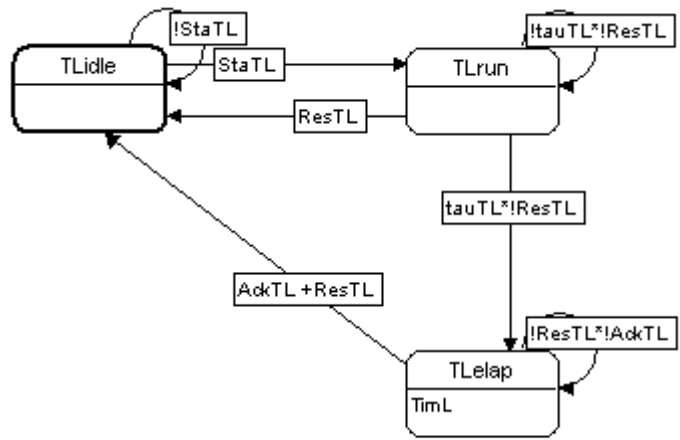

**Fig. 6. The model of the TIMER TL**

### 3.2.3 CSM model of the TIMER TL

TIMER TL, for indicating 'long' time intervals (Fig. 6.), has its states similar to the ones of TIMER TS. However, as it was mentioned before, there are two new ways of concluding the timer's activity while TIMER TS, once started, performed the fixed sequence of states spontaneously. Now, the CONTROLLER can set the timer back to its initial state either by the use of the *AckTL* signal (only from the *TLelap* state) or reset it instantly from *TLelap* as well as from *TLrun,* using the *ResTL* signal.

## 3.3. Reachability Graph of a system

The graph of reachable system states, obtained with the appropriate module of the COSMA 2.0, is shown in Fig. 7. It is again a CSM, with system states being the vectors of states of components and transition formulas as easily interpretable as in the case of individual components. The inspection of this Reachability Graph 'by naked eye' shows that the system performs properly.

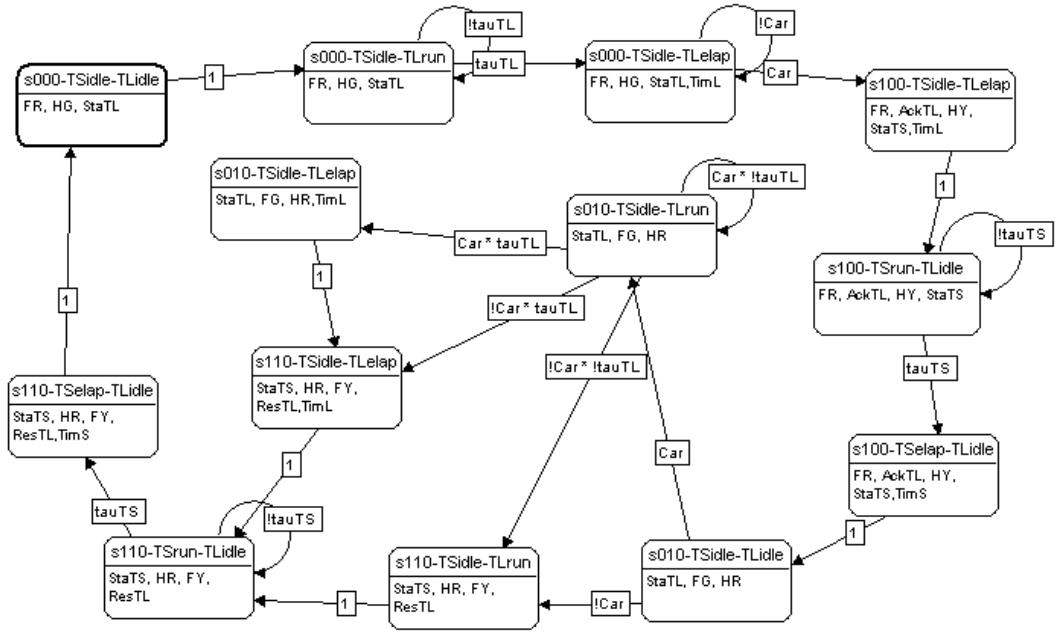

**Fig. 7. Reachability Graph of the TLC system**

The above graph is shown just as the illustration of the idea underlying the CSM model and the COSMA methodology. It has as few as 13 states[6] so that it can be actually drawn or printed and analyzed 'by naked eye'. In practical cases, where number of states can be of order of $10^{20} - 10^{50}$ or even more [5], inspection of the RG has to be done 'automatically', i.e. by the algorithm for the evaluation of special formulas which represent *requirements* to the system's behavior. This issue is discussed in the next section.

## 4. Temporal model checking of a system

The controller (consisting of main automaton and automata of timeout devices) was checked for formal correctness. A temporal model checker TempoRG, which is an element of COSMA 2.0 environment, was applied for verification. First, some questions asked originally by Ramachandran [19], but translated to temporal sentences, were asked. Questions by Ramachandran have the form checking the truth of sentences expressed in natural language, for example:

---

[6] Note, however, that it is only 13 states out of 36 elements of the Cartesian product of three components, having 4, 3 and 3 states, respectively. The remaining ones are *unreachable* states and they have been 'cut off' by the algorithm based on the CSM theory.

```
If there is a car on the side road and the long timeout
occurs, the highway light changes to yellow.                    (1)
```

The original correctness conditions will be referred to by a number of "strategy" followed by a number of condition. A translation to temporal formulas is as follows:

$\Box ((HG * Car * TimL) \Rightarrow (\circ HY))$ (1)
$\Box ((HY * TimS) \Rightarrow (\circ (HR * FG)))$
$\Box ((FG * (!Car)) \Rightarrow (\circ FY))$
$\Box ((FG * TimL) \Rightarrow (\circ FY))$
$\Box ((FY * TimS) \Rightarrow (\circ (HG * FR)))$

Two Ramachandran's conditions were not checked: initial condition (it is static, non-temporal condition) and the condition:

```
If there is no short timeout, then the Highway light changes to red
and the farm light changes to yellow
```

because the authors of this paper do not understand the meaning of the test.

When asked, all the tests returned the value *true*.

Then, next test was performed to check what would occur if the stream of farmer cars never ends (it is modeled by signal *Car* lasting infinitely). External signals are assumed to be fair in CSM (any combination of signals is possible in any moment, including no signal). Therefore, an additional automaton was added to a system. It is a two-state automaton with terminal state generating signal *Car* (Fig. 8). Again, all questions are answered positively.

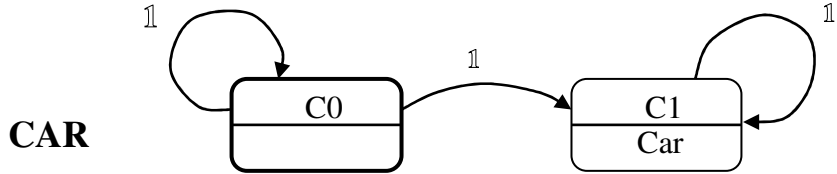

**Fig. 8. A *CAR* automaton modeling unfair situation (signal *Car* lasting infinitely)**

The formulas contain a "*next-step*" operator, as they are converted from original Ramachandran tests (Appendix B). We suggest that the formulas should not take next step, as it depends on "granularity" of actions considered. The better testing is based on a "future" behavior assuming a state reached. Such questions have a form:

$\Box ((HG * Car * TimL) \Rightarrow (\Diamond HY))$ (1)
$\Box ((HY * TimS) \Rightarrow (\Diamond (HR * FG)))$
$\Box ((FG * (!Car)) \Rightarrow (\Diamond FY))$
$\Box ((FG * TimL) \Rightarrow (\Diamond FY))$
$\Box ((FY * TimS) \Rightarrow (\Diamond (HG * FR)))$

All these questions are answered positively, as before.

**5. Guidelines for the generation of VHDL code from CSM specification**

After a "manual" synthesis of a TLC controller, we have tried to obtain a hardware description code in VHDL [14] automatically. If has followed some attempts to specify code generation rules for CSM automata [15,16,17]. The following rules were assumed for VHDL code generation:

i. External signals (input alphabet of an automaton) and generated signals (output alphabet of an automaton) are modeled as hardware signal lines:

   Car : in BIT;
   TimL : in BIT;
   TimS : in BIT;
   StaTL : out BIT;
   AckTL : out BIT;

ii. States of an automaton are coded on a vector of boolean variables (to be synthesized as flip-flops) of desired dimension. Although two flip-flops are enough to code states of four-state CONTROLLER automaton, three variables are used to fit the previous encoding in this paper. Three vectors are used for preparing moving to new state while staying in a given state (*current_state*, *state* and *newstate*). Initial values of state vectors are assigned as constants.

   variable current_state : BIT_VECTOR (2 downto 0) :='000';
   variable state; newstate : BIT_VECTOR (2 downto 0) :='000';

iii. For every generated signal, an additional variable is used that stores its "prepared" value (to be latched after a clock signal comes):

   variable newHG: bit;

iv. Every automaton is modeled as a process in VHDL:

   architecture TLC of TLC is
   begin
   traffic:process
   ...
   end process traffic;
   end TLC;

v. Arcs leading out of a state are modeled as conditional statement (if there is more than one arc) or compound statement (if there is only one arc) preparing new values of state vectors and generated signals:

```
                when "000" => if (Cars='1') and (TimL='1') then
                                    newstate:='100';
                                    newSTL:='0'; newRTL:='0'; newATL:='1';
                                    newSTS:='1';
                                    newHG:='0'; newHY:='1'; nweHR:='0';
                                    newFG:='0'; newFY:='0'; newFR:='1';
                              else
                                    newstate:='000';
                                    newSTL:='1'; newRTL:='0'; newATL:=0;
                                    newSTS:='0';
                                    newHG:='0'; newHY:='1'; nweHR:='0';
                                    newFG:='0'; newFY:='0'; newFR:='1';
                              end if;
```

vi. At the end of every cycle of infinite loop of an automaton, new values are assigned to state vector and to all generated signals. Then, a delay of 10ns is applied for analogous circuitry (as the environment of VHDL specification is synchronous).

```
        state<=newstate;
        StaTL<=newSTL;
        AckTL<=newATL;
        HG<=newHG;
        HY<=newHY;
        wait for 10 ns;
```

The presented rules are applied in slightly different ways in most of VHDL CAD environments, i.e. the environment contains an FSM graphic editor, which allows to define an automaton as graph, and to store it in a text file. Next, the file is used to generate a VHDL description of the automaton. The code generated by distinct tools differ. The differences arise from the fact that synthesisable subsets of VHDL accepted by the tools do not match.

We have used Active-VHDL environment from ALDEC Inc. to generate VHDL code from our automata. The results differ from rules presented above by:
- presence of an explicit clock signal - the environment assumes synchronous implementation of circuits;
- symbolic names of states instead of bit-vectors; the environment assumes that the VHDL code will be passed to a logic synthesis tool to obtain an optimal encoding.

We argue that the COSMA environment can be efficiently used for analysis and validation of control circuits which are planed to be synthesised in hardware. The translation of automata specification in CSM to a VHDL source code is simple, and we are sure that the VHDL code generated according to above outlined rules will be accepted by the environment for hardware synthesis. Moreover we are sure that the resulting circuit is correct with respect to the questions we have asked and checked in COSMA.

**6. Concluding remarks**

In addition to the tutorial values of the discussion presented in the paper, several following conclusions are worthy to be emphasized:

- Specification of the behavior in terms of the CSM model is easily understandable and close to the common intuition,
- On the other hand, CSM model is a formal one and supports the formal verification of system's behavior,
- Software modules making the COSMA environment perform properly, at least in the scope of functions required for this type of analysis (edition of components' graphs, calculating the Reachability Graph of a system, temporal model checking),
- The CSM specification facilitates the generation of VHDL code (using a subset of VHDL language),

Therefore, the CSM model and the methodology based upon the COSMA environment can be effectively used at system level synthesis of digital circuits.